# Toward dynamical crowd control to prevent hazardous situations

Tomoichi Takahashi
Department of Information Engineering
Meijo University
Nagoya, Japan

1. Introduction

It is common for large crowds to gather to attend games, exhibitions, political rallies, and other events. Thus, careful designs and operational plans are made to ensure the safe, secure, and efficient movement of people in these crowded environments. However, the congestion created by large crowds has resulted in hazardous incidents across the world (e.g., those during the Hajj pilgrimage, gatherings at Mecca, and the Akashi fire festival) [1][2].

In 2011, the Tokyo Fire Department conducted a study on fire-prevention plans at large subway terminal stations [3]. The study surveyed 400 people who responded to questions such as "What points do you feel anxious about in emergency situations?" and "How do you behave when a fire occurs?" The survey revealed that the majority of people in unfamiliar places follow the guidance provided by public announcement (PA) systems and the people around them.

Figure 1 displays photos captured at Tokyo rail terminals after a snowfall in January 2018. People returned home early to avoid potential troubles resulting from the snowfall. The number of people present at the terminals was larger than the allowable transportation capacity. People consequently filled the terminals, causing the railroad companies to restrict entrance. However, people remained disciplined under this duress for several hours in part because they were able to ascertain the causes of the congestion and could access information through their cellular telephones.

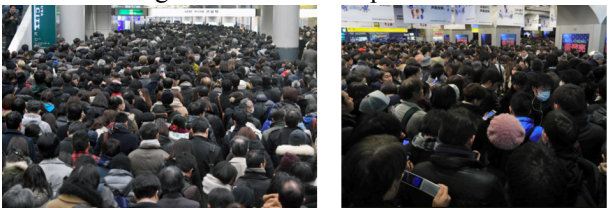

Figure 1. Crowded terminals (2018. Jan.22)

Developments in information technology (IT) can provide new means to disseminate public information, thus changing human behavior in situations of danger and duress. In this paper, we propose a crowd control and evacuation guidance (CCEG) management system using digital promotional signage (PS) to demonstrate the effects of crowd control via simulations.

2. Providing information and guidance

2.1 Preliminary studies

Figure 2 depicts signs often seen in public spaces across the world. The first image shows signs indicating the direction in which emergency exits can be found. The second image shows a wall-mounted evacuation plan and map. The exit sign (ES) and map are considered to be good methods for prompting evacuations. However, people seem to travel without paying attention to these signs unless unexpected events occur and the guidance contents are fixed.

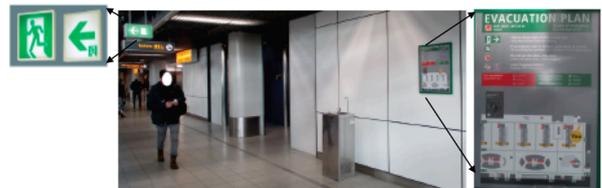

Figure 2. Evacuation guidance media at a terminal

We conducted a survey to determine whether people sufficiently notice signs during everyday life. This study involved two questionnaire surveys (i.e., test A and test B), which were conducted at a subway mall in 2016 (Figure 3). The subway mall is located underground and contains fourteen exits leading to the surface and adjacent buildings. The test subjects were different groups of students at our university. We asked these subjects to walk normally in the mall along the route displayed in Figure 3 on different days. At the end of the route, the subjects were asked to recall the number and locations of any ESs or digital PS they saw.

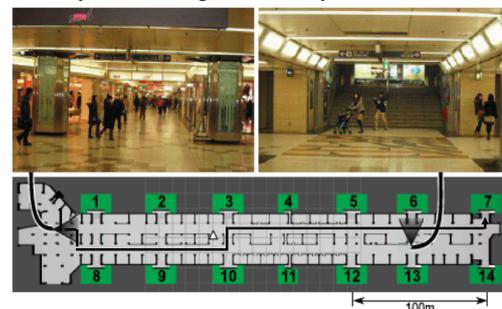

Figure 3. Underground shopping mall (black line indicates the route traveled by participants)

In test A, participants were asked to retrace their paths and attentively count the number of ESs and digital PS. This exercise corresponded to real emergency situations in which evacuees would search for guidance. Such an event was staged during test B; so seven additional PS locations were included.

Table 1 indicates the number of signs that were spotted along the route shown in Figure 3. The results demonstrate the following:
- Participants noticed a maximum of 28% of the ES locations when they traversed the route as usual, but detected nearly three times as many when asked to pay attention.
- There was little difference in the perception rates between the proportion of ESs and PS noticed.
- People unconsciously noticed a higher proportion of PS during the unexpected staged event.

This indicates that PS functions as ESs during emergency situations. Thus, an increased number of signs can guide more people to proper evacuation routes.

Table 1. Survey on ES awareness

| Test | Sign | Number of signs | Number and rate awareness | |
|---|---|---|---|---|
| | | | unaware(forth) | Attentive(back) |
| A | ES | 36 | 9.3 (26%) | 30.8 (86%) |
| | PS | 5 | 1.4 (28%) | 4.0 (80%) |
| B | ES | 36 | 6.0 (16%) | - |
| | PS | 12* | 9.0 (75%) | - |

(*Seven PSs were added for an unexpected event)

2.2 New guidance systems proposals

PSs have recently been used for promotions and advertising in public spaces. Figure 4 shows an image of our CCEG management system. The first figure shows everyday use in which PS is used to display promotional contents. The second figure shows our CCEG concept in which the PS is used to display contents suitable for emergencies. In this concept, the role of the PS changes to that of an ES. That is, the PS displays messages intended to guide people to safety. To accomplish this, data from video camera feeds are sent to crisis-management centers, proper evacuation routes are planned for emergencies, and PS messages are then displayed to provide proper information based on these plans.

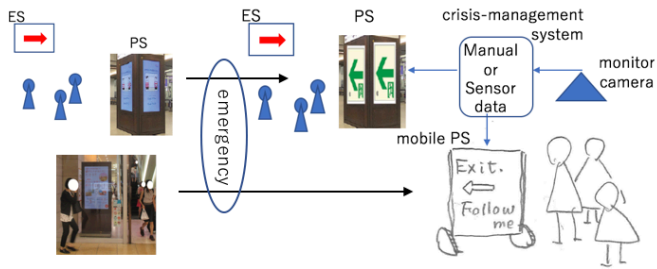

Figure 4. Crowd control management system using PS

3. Validation via agent-based simulations

3.1 Simulation scenarios

It is necessary to show the effectiveness of the CCEG before it is practically applied. The agent-based social simulation (ABSS) is an appropriate testing method for this. The CCEG's effectiveness was validated by simulating crowd behavior at the subway mall through the TENDENKO ABSS system (Figure 3). Agents perceived the surrounding data and selected an action from a prescribed set at every step of the simulation [4].

An alarm was first sounded through the PA system. A verbal warning indicated the following: "Fire near exit number one; evacuate from exit number seven." This message was sent to all agents, who were assumed to have begun following the instructions. However, behavior becomes unpredictable during emergencies. Individual movement creates congestion, and evacuation efficiency is reduced. To increase efficiency, PS is used to display proper directions and information about evacuation routes according to manuals and sensor data. Agents within 10$m$ of the PS can view these messages.

3.2 Simulation validation

The crowd evacuation simulation is a type of social simulation. The validation of these simulation tools has been a highly important issue, and a number of questions have been raised as a result of the associated testing methods. The following questions were suggested according to the quantitative/qualitative attributes of behavioral uncertainty, which are essential during ABSS experiments [5]:

Q1. How do we ascertain whether a tool is sufficiently accurate?
Q2. Which and how many tests should be performed to assess the accuracy of the model's predictions?
Q3. Who should perform the tests (e.g., the model's developers, users, or a third party)?

These questions should be answered for users who will employ the simulation results when making practical decisions.

3.3 Simulation parameters

Personal behavior was simulated according to the following three changing parameters:

$p$ is the rate at which agents follow PS guidance. Reports from the Great West Japan Earthquake of 2011 indicated that some agents evacuated immediately, but others did not [6]. This parameter is indicative of how human behavior differs according to the individual and their surrounding environment.

$n$ is the number of people present at the mall. They are uniformly distributed throughout this location.

$s$ is the number of PS locations used as ESs during emergencies. PS positions were selected to form test B.

Simulations were executed in which $n = \{1000, 4000, 7000, 10000\}$, $p = \{30\%, 50\%, 70\%, 100\%\}$, and $s = \{4, 6, 8\}$. The population density of $n = 4000$ is equal to one person per 1$m^2$. This number corresponds to crowd status during commuting hours. Figure 1 depicts a crowd congestion status of $n = 10000$.

Two message display modes were compared:

Default mode: At the time of the first PA announcement, the PS display was changed to "Go to the nearest exit. The exit number is X."

Congestion mode: When congestion occurred, the PS display changed to "Go to the nearest exit; the number is X." Two

sign-change policies were tested to ascertain the effects of individuals detecting the congestion status:
policy 1 employs population density per square meter;
policy 2 uses the changes in population density

## 4. Simulation and decision-making analyses
### 4.1 Effects of increasing the amount of PS

Figures 5 and 6 show the simulation results for n = 4000. The first and second charts in Figure 5 correspond to simulations containing no signs and a default simulation mode of $p$=30%. The vertical axis in each graph represents the evacuation rate (i.e., the rate of individual evacuation). The horizontal axis in each graph represents simulation time. The graphs show that evacuation rates improved according to the amount of PS, even with low $p$.

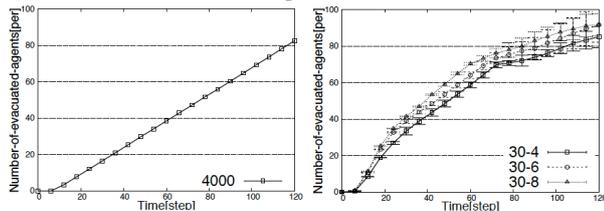

Figure 5. Improved evacuation rate ($n$=4000 and $s$=8) (Legend numbers are $p$-$s$, e.g., 30-4 corresponds to a simulation of $p$=30% with four PS locations.)

### 4.2 The effects of introducing congestion

The first and second charts in Figure 6 correspond to the default and congestion modes (policy 1), respectively. Both graphs indicate that the evacuation rates were improved with increased $p$. Higher evacuation rates are shown on the right-side chart compared to the corresponding rates on the left-side chart. This indicates that adaptive evacuation guidance improved the state of congestion, and consequently the evacuation rate.

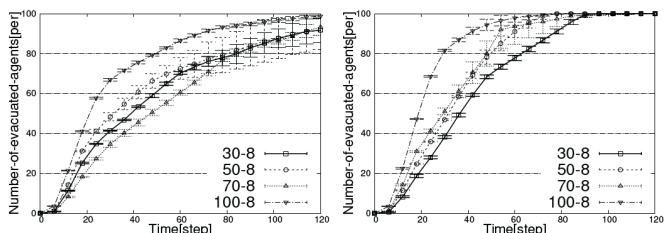

Figure 6. Improved evacuation rate ($n$=4000 and $s$=8)

### 4.3 Effects of congestion status detection

Figure 7 shows the results of comparing the two policies for $p$ = 0.7. The first and second charts depict $n$=4000 and 10000. The p1 graph in the first chart corresponds to the legend indicating 70-8 in the second chart of Figure 5. Policy 2 indicates better results as steps advance for both $n$=4000 and 10000. This indicates that the ways of evacuation detection improves evacuation rates.

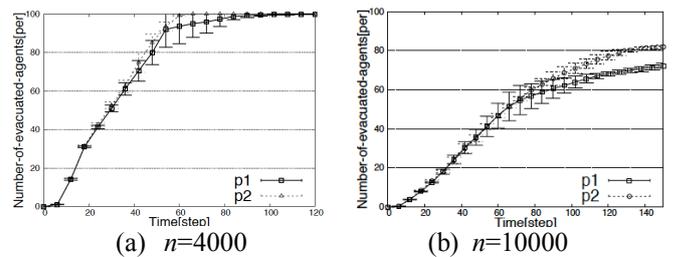

Fig. 7 Evacuation rate ($p$=0.7 and $s$=8). p1 and p2 in the legends correspond to policy1 and policy2, respectively.

## 5. Discussion and summary

Technological advancements influence how we access information and provide data. New media formats will affect human behavior during emergencies in ways that we can only assume at present. This means that there is little data to prove the effects of the new technology.

This study proposed a CCEG management system to prevent hazardous crowd situations. This system adopted the PS media format as a new disaster prevention method for future disasters. The effects of PS on evacuation rates were simulated through ABSS and disaster simulations should be designed to ascertain the behaviors of study targets in addition to the features of the field before the results are used during policy creation.

These simulations indicated that an increased number of signs prompted evacuations and that sign contents affected evacuation behaviors. Such validating approaches involving field simulations designed to study human behaviors will be important for assessing new disaster-prevention technologies.

## 6. Acknowledgements

The authors would like to thank Katsuki Ichinose for conducting the simulations. This work was supported by JSPS KAKENHI Grant Number 16K01291.

## REFERENCES

[1] C. Tunasar, "Analytics driven master planning for mecca: Increasing the capacity while maintaining the spiritual context of hajj pilgrimage". 2013 Winter Simulations Conference (WSC) (pp. 241-251). IEEE

[2] Akashi City: Fire festival (in japanese), http://www.city.akashi.lg.jp/anzen/anshin/bosai/kikikanri/jikochosa/index.html

[3] Tokyo Fire Department. Fire evacuation simulation at big terminal station (in japanese), http://www.tfd.metro.tokyo.jp/hp-yobouka/fukugouterminalanzen/

[4] M. Okaya, T. Takahashi: Effect of guidance information and human relations among agents on crowd evacuation behavior, 6th International Conference on Pedestrian and Evacuation Dynamics Programme (PED2012).47-48, 2012, July

[5] Averill, J. D., et.al. (2005). Occupant behavior, egress, and emergency communications (Draft). U.S. Department of Commerce, National Institute of Standards and Technology. Gaithersburg, MD: National Institute of Standards and Technology. Retrieved from https://ws680.nist.gov/publication/get_pdf.cfm?pub_id= 909233

[6] Cabinet Office Government of Japan. Prevention Disaster Conference, the Great West Japan Earthquake and Tsunami. Report on evacuation behavior of people. http://www.bousai.go.jp/kaigirep/chousakai/tohokukyokun/7/index.html